\documentclass[final,cmfonts]{aipproc}

\layoutstyle{8x11double}

\newcommand{\al}{\alpha}
\newcommand{\be}{\begin{equation}}
\newcommand{\ee}{\end{equation}}
\newcommand{\bdm}{\begin{displaymath}}
\newcommand{\edm}{\end{displaymath}}
\newcommand{\bea}{\begin{eqnarray}}
\newcommand{\eea}{\end{eqnarray}}

\newcommand{\Sc}{{\cal S}}
\newcommand{\Om}{\Omega}

\newcommand{\Md}{\dot{M}}
\newcommand{\Mda}{\langle\dot{M}\rangle}

\newcommand{\tb}{{\cal T}_B}

\begin{document}

\title{A Timing Signature of Gravitational Radiation from LMXB Neutron Stars}

\author{Robert V. Wagoner}{address={Dept. of Physics and KIPAC, Stanford University, Stanford, CA 94305-4060},email={wagoner@stanford.edu}}

\begin{abstract}

The coupled evolution of the spin frequency, core temperature, and r--mode amplitude of an accreting neutron star is calculated. We focus on those conditions that can produce persistent gravitational radiation from the r-mode. During X-ray quiescent phases of transient LMXBs, one may be able to identify the constant contribution of the gravitational wave emission to the spindown rate. Another signature is the r--mode contribution to the heating.  

\end{abstract}

\maketitle

There is now strong evidence \citep{cmm03} that the frequency $f=1/P$ of burst oscillations in accreting low-mass X-ray binaries (LMXBs) asymptotes to the spin frequency of the neutron star, as separately measured in the coherent millisecond X-ray pulsar LMXBs.  Moreover, the burst oscillations are often also sufficiently coherent to allow a determination of the spindown (or spinup) rate $P^{-1}dP/dt$. The distribution of spin periods indicates that gravitational radiation may be limiting the accretion--driven spinup of these neutron stars \citep{bil98,wag84,cmm03}. Here we investigate the possibility that steady spindown produced by the emission of gravitational radiation via an excited r--mode might be detectable during the quiescent phases of these X-ray transients.         

\subsection{Governing relations}

Details of the physical conditions and evolution of neutron stars that have been spun up by accretion to a state in which gravitational radiation can be emitted from an r--mode, and references to its earlier development, are provided by \citet{wag02}. The dominant ($l=m=2$) r--mode (velocity perturbations $\delta v$ driven by the Coriolis force) is characterized by an amplitude $\al\sim \delta v(r=R)/\Om R$, where $\Om=2\pi/P$ and $R$ are the angular velocity and radius of the neutron star. With $J_*(M,\Om)$ the angular momentum of the equilibrium neutron star, its total angular momentum is $J = J_* + (1-K_j)J_c$, where the canonical angular momentum is $J_c = -K_c\al^2 J_*$. (These constants $K_{(\;)}\sim 1$.)  

The angular momentum perturbation obeys the relation
\begin{equation}
   dJ_c/dt = 2J_c[(F_g(M,\Om)-F_v(M,\Om,T)] \; . \label{dJcdt} 
\end{equation}
The gravitational radiation growth rate is $F_g = \tau_{gr}^{-1}(\Om/\Om_c)^6$, with $\tau_{gr}\sim 3$ s for the neutron star model adopted and $\Om_c\equiv(\pi G\langle\rho\rangle)^{1/2}$. In what follows, we shall neglect the dependence on the slowly increasing mass $M$.

The viscous damping rate is 
\[ F_v \cong F_{bl}(\Sc_n,\Sc_s,B,\Om,T) + F_{hb}(T_h,\Om,T) \; , \]
where $T$ is the core temperature, and $B$ is the magnetic field strength in the core--crust boundary layer. We have modified the viscous and magnetic boundary layer damping rate ($F_{bl}$) of \citet{km03}, with the splippage factors \cite{lu01} $\Sc$ for the normal and superfluid components. The hyperon bulk viscosity damping rate ($F_{hb}$) due to $n+n\rightleftharpoons  n+\Lambda, \; p+\Sigma^-$, etc. employs results of \citet{lo02} and \citet{hly02}. (But very recently \citet{vdd03} obtained a smaller rate.)

Conservation of total angular momentum then requires that
\[ dJ/dt = 2J_c F_g + j_a\Md(t) - \tb(\Md) \; . \]
The first (negative) term is due to gravitational radiation, the second to accretion [at the rate $\Md(t)$], and the third is the magnetic torque (e.g.~\citep{rfs03}) of the disk on the star.

From these relations, we can obtain the spin evolution equation: 
\begin{eqnarray}
 \left(\frac{I_*}{J_*}\right)\frac{d\Om}{dt} &=& -2[K_jF_g+(1-K_j)F_v]K_c\al^2 \nonumber \\                                             &+& \frac{(j_a-j_*)}{J_*}\Md(t) - \frac{\tb}{J_*} \; , \label{dOdt}
\end{eqnarray}  
where $I_*(M,\Om)=\partial J_*/\partial\Om$ and $j_*(M,\Om)=\partial J_*/\partial M$.

Finally, thermal energy conservation for the star gives
\begin{eqnarray}
\int{\partial T\over\partial t}c_vdV &\equiv& C(T){dT\over dt} \cong 2\tilde{E}_c F_v(\Om,T)  \nonumber \\
&+& K_n\Mda_n c^2 - L_\nu(T) \; , \label{dTdt}
\end{eqnarray}
where the r--mode viscous heating is proportional to the rotating frame canonical energy ${\tilde E}_c = (K_c/3)J_*\Om \al^2$.  We have assumed that the thermal conductivity timescales are short enough to allow the use of a single spatially-averaged core temperature $T$. The inner crust nuclear heating is driven by an accretion rate $\Mda_n$, averaged over the nuclear time scale $\tau_n\sim 1$ year \citep{wgv02}. The constant $K_n\approx 1\times 10^{-3}$.

Comparison of observations of thermal emission from isolated neutron stars with computed cooling histories has led \citet{kyg02} to propose that at the core temperatures of interest here ($T\sim 3\times 10^8$ K) the core (triplet) neutrons are normal, while the core (singlet) protons and inner crust (singlet) neutrons are superfluid. The neutrino lumionosity that we employ \citep{wag02} is then due to direct Urca, modified Urca, electron--ion and neutron--neutron neutrino bremsstrahlung, and Cooper pairing of (inner crust) neutrons.

For typical values of $\Om$, $T$, and $\Mda < \Md_{Edd}\sim 3\times 10^{-8}M_{\odot}/$yr (with $j_a\sim\sqrt{GMR}$), we obtain three key time scales (gravitational radiation, cooling, and accretion): $\tau_g \equiv 1/F_g \sim 10^3\mbox{ s}$ , 
\[ \tau_c \equiv \frac{C(T)T}{L_\nu(T)} \sim 10^3\mbox{ yr} \; ,\quad 
   \tau_a \equiv \frac{J_*}{j_a\Mda_a} > 10^7\mbox{ yr} \; .\]
These time scales are seen to be relaxation times of our evolution equations (\ref{dJcdt}), (\ref{dTdt}), and (\ref{dOdt}), respectively. Here the accretion rate is averaged over the time scale $\tau_a$.

\subsection{Evolution}

We are mainly interested in the evolution of neutron stars after they have been spun up to the point where equation (\ref{dJcdt}) vanishes: $F_g(\Om_{in})=F_v(\Om_{in},T_{in})$. 
This equality defines our initial state, where the perturbation can begin to grow.
The initial temperature $T_{in}$ is determined by the vanishing of equation (\ref{dTdt}), with the nuclear heating balanced by the neutrino emission.

In contrast to the initial state, the equilibrium state is defined by the vanishing of the evolution equation (\ref{dOdt}), in addition to the evolution equations (\ref{dJcdt}) and (\ref{dTdt}). The equilibrium amplitude is then given by $\al_{eq}=[\tau_g/(2K_c\tau_a)]^{1/2} \sim (10^{-7}-10^{-5})$, assuming that the magnetic torque is negligible.

In Fig.~1, we show the critical curve defined by $F_g(\Om)=F_v(\Om,T)$, for a somewhat optimistic choice of damping rate parameters. On it are indicated the initial and equilibrium states for two choices of (time averaged) mass accretion rate. The linear perturbation analysis of \citet{whl01} shows that stability of the equilibrium state requires that the slope $d\Om/dT > 0$. 

\begin{figure}[htb]
  \resizebox{\columnwidth}{!}{\includegraphics{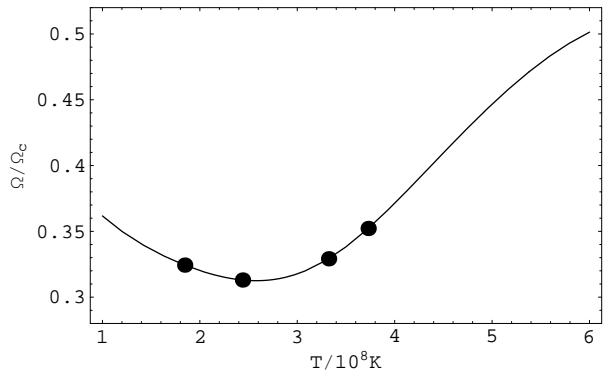}}
  \caption{The relation between dimensionless angular velocity $\Om/\Om_c$ ($\approx 2/3$ at the shedding limit) and temperature $T_8$ on the critical curve defined by $F_g(\Om)=F_v(\Om,T)$. The model chosen has a normal fluid hyperon damping rate 10 times greater than that of \citet{lo02}, a hyperon superfluid transition temperature $T_h=1\times 10^9$ K, core--crust slippage factors $\Sc_n=\Sc_s=0.2$, and boundary-layer magnetic field $B<10^9$ G. Also shown are the initial and equilibrium states for (a) $\Mda_a=\Md_{Edd}/300$: $T_8(in)=1.84$, $T_8(eq)=2.44$ and for (b) $\Mda_a=\Md_{Edd}/3$: $T_8(in)=3.32$, $T_8(eq)=3.73$.}
\end{figure}

It can be shown that the evolution from the initial state is also controlled by the sign of the slope $d\Om/dT$ of the critical curve:

\noindent$\bullet$ If the slope is negative [as in Fig.~1, case (a)], there will be a thermal runaway ($dT/dt >0$, $d\Om/dt\approx 0$) with a growth rate $K_r^{1/2}\al_{eq}/\tau_g\sim 1/$yr that is of the same magnitude as found by \citet{lev99}. Here $K_r\sim 10^5$ is the ratio of rotational to thermal energy in the star. The initial phase of the runaway occurs at roughly constant angular velocity. The later evolution is uncertain, but the neutron star may again reach the critical curve, where its slope could be positive.

\noindent$\bullet$ If the slope is close to zero, there will initially be overstable oscillations of the type found by \citet{whl01}. 

\noindent$\bullet$ If the slope is positive [as in Fig.~1, case (b)], the oscillations of the growing amplitude are damped out on a timescale $\sim\tau_c$, after which it slowly increases toward its equilibrium value, as shown in Fig.~2. The evolution of $\Om$ and $T$ is similar, with the time required to reach the equilibrium state given by 
\[ \Delta t\approx [(\Om_{eq}-\Om_{in})/\Om]\tau_a \; . \]
If the magnetic torque term in equation (\ref{dOdt}) were important, $\tau_a$ could be appropriately modified. After accretion ceases, the star spins down along the critical curve. 

\begin{figure}[htb]
  \resizebox{\columnwidth}{!}{\includegraphics{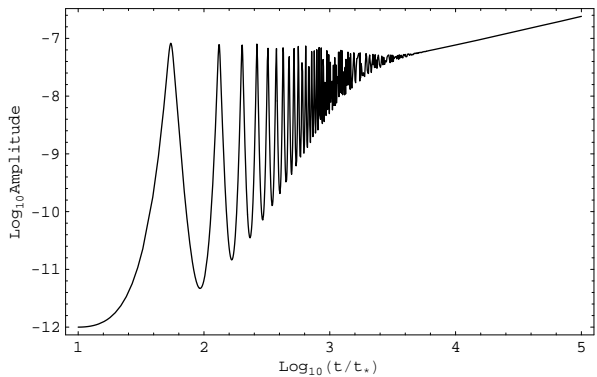}}
  \caption{The early evolution of the r-mode amplitude $\al$, for $\Md=\Md_{Edd}/3$. The initial amplitude (during spinup to the critical curve) was chosen to be $\al_0=10^{-12}$, and $t_*=1$ yr. When equilibrium is reached, $\al_{eq}= 3.60\times 10^{-6}$.}
\end{figure}

The ratio of the spindown rate to spinup rate is
\begin{equation}   
\frac{P^{-1}dP/dt(\Md\ll\Mda_a)}{P^{-1}dP/dt(\Md=\Mda_a)} 
\approx -\left(\frac{\Om}{\Om_{eq}}\right)^6\left(\frac{\al}{\al_{eq}}\right)^2 \; . \label{sp}
\end{equation}   
We also find that on the critical curve, the ratio of r--mode heating to accretion-induced nuclear heating in equation (\ref{dTdt}) is
\begin{equation}
\frac{2\tilde{E}_c F_g}{K_n\Mda_n c^2} \approx 10\frac{\Mda_n(eq)}{\Mda_n} \left(\frac{\Om}{\Om_{eq}}\right)^8\left(\frac{\al}{\al_{eq}}\right)^2   \; . \label{heat}
\end{equation}

In all cases, if the neutron star reaches a positive slope section of the critical curve in any way, equations (\ref{dJcdt}) and (\ref{dTdt}) quickly relax to equilibrium. So further evolution occurs at a rate governed by equation (\ref{dOdt}) (whose first term is now $-2K_cF_g\al^2$), with steady gravitational radiation generated by the r--mode amplitude given by the vanishing of equation (\ref{dTdt}) (with $F_g=F_v$):
\begin{equation}   
\al^2 = (L_\nu-K_n\Mda_n c^2)/[(2K_c/3)J_*\Om F_g] \; . \label{al2}
\end{equation}
This remains true as long as $d\Om/dT>0$. After accretion ceases, the star will spin down until this slope vanishes \citep{rb03}. Then the star will cool further at constant $\Om$, no longer emitting gravitational waves ($\al$ negligible). It then can become a millisecond radio pulsar \citep{rb03}. 

After $T$ increases slightly above $T_{in}$, the neutrino luminosity dominates the nuclear heating (for constant $\Mda_n$) in equation (\ref{al2}). We plot the resulting amplitude at any point on the positive slope section of the critical curve in Fig.~3. Using this amplitude, we show in Fig.~4 the spindown rate when the accretion and magnetic torques are smaller than that due to gravitational radiation in equation (\ref{dOdt}). 

\begin{figure}[htb]
  \resizebox{\columnwidth}{!}{\includegraphics{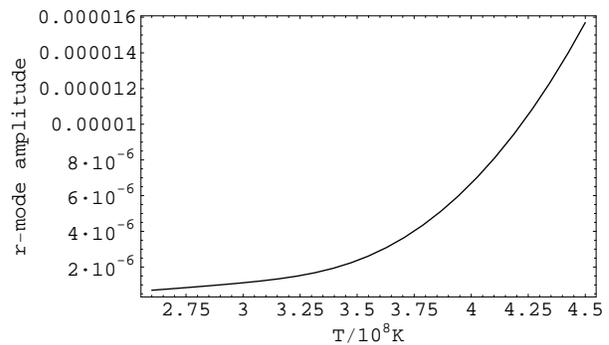}}
  \caption{The r--mode amplitude $\al$ at any point on the stable section of the critical curve where $L_\nu\gg K_n\Mda_n c^2$.}
\end{figure}

\begin{figure}[htb]
  \resizebox{\columnwidth}{!}{\includegraphics{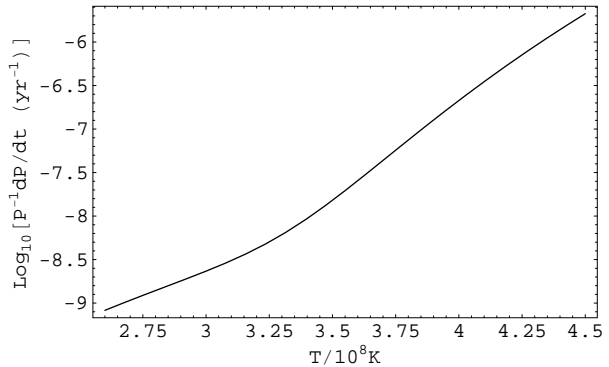}}
  \caption{The spindown rate produced by gravitational radiation when the neutron star is on the stable section of the critical curve.}
\end{figure}

Observations of transient LMXBs during their quiescent states (when $\Md(t)\ll\Mda$) provide the opportunity to detect two signatures that the neutron star is emitting gravitational radiation at the rate predicted if it has evolved to the stable portion of the critical curve. In order that such a positive slope portion exist, it appears that the neutron star core must contain hyperons with a moderate superfluid transition temperature ($\sim 10^9$ K) and neutrons with a low superfluid transition temperature ($\sim 10^8$ K) \citep{wag02}.  

The first signature was pointed out by \citet{bu00}, who showed that observations of the X-ray luminosity during quiescence can place interesting limits on the amount of r--mode heating. We have also estimated the photon luminosity, and obtained results similar to theirs for a `normal core'. They assumed that the star was in the equilibrium state. Note however from equation (\ref{heat}) that the mode heating will often be less than they obtained during spinup to or spindown from this state. Within the uncertainties of the mass and neutrino luminosities consistent with observations of isolated neutron stars  \citep{ylp03}, comparison with observations of the four quiescent sources with large spin frequencies allows the presence of the predicted level of r--mode heating. However, this extra source of heating is certainly not yet required \citep{ylp03}. 

Here we focus on the second signature. When the accretion torque becomes negligible at sufficiently small values of $\Md(t)$, the constant gravitational wave torque provides a fixed lower limit to the spindown rate, shown in Fig.~4. If the contribution of magnetic torque can be identified from its dependence on $\Md(t)$ and removed, a constant residual will provide evidence that the neutron star is emitting gravitational radiation at a significant rate. The transient XTE J0929-314 was observed to have an average spindown rate $P^{-1}dP/dt\approx 1.6\times 10^{-8}$ per year \citep{gcm02}, comparable to that in Fig.~4 for a core temperature $T\approx 3.5\times 10^8$ K. The transient SAX J1808.4-3658 was observed to spin down at about the same rate after an outburst in 2002, but spun up by a larger amount during its 1998 outburst \citep{mgc03}. As more data accumulate, it may be possible to extract more information about the contributions to spindown in the candidate (high spin frequency) neutron star transients. 

If Sco X-1 has been spun up by accretion to a stable equilibrium state (in which gravitational-wave flux is proportional to X-ray flux) \citep{wag84}, it should be detectable by Advanced LIGO. When signal recycling (narrow-banding) is employed, a few additional LMXB's may also be detectable \citep{bil03,ct02}. A major uncertainty is the actual value of the (very) long--term average $\Mda_a$, as inferred from the X-ray flux. If the neutron star is not in the equilibrium state, but is on the stable portion of the critical curve, the gravitational-wave amplitude (metric perturbation) will scale as $h\propto\Om^3\al$. For the dominant r--mode, the gravitational wave frequency is $f_{gw}\cong 4/(3P)=2\Om/(3\pi)$.  

\begin{theacknowledgments}
This work was supported in part by NSF grant PHY--0070935.
\end{theacknowledgments}

\bibliographystyle{aipproc}

\end{document}